\documentclass[a4paper]{jpconf}
\usepackage{graphicx}
\usepackage{epsfig}
\begin{document}
\title{${\cal {B}}(D_s^+\to\ell^+\nu$) and the Decay Constant
$f_{D_s^+}$}

\author{Sheldon Stone}

\address{Physics Department, Syracuse University, Syracuse N.Y. 13244, USA}

\ead{stone@physics.syr.edu}

\begin{abstract}
I report final CLEO-c results on the purely leptonic decays of the
$D_s^+\to \ell^+\nu$, for the cases when $\ell^+$ is a $\mu^+$ or
$\tau^+$, when it decays into $\pi^+\overline{\nu}$ using 314
pb$^{-1}$ of data at 4.170 GeV. I also include preliminary results
from the $\tau^+\to e^+\nu \overline{\nu}$ channel using 195
pb$^{-1}$. Combining both we measure $f_{D_s}=275\pm 10 \pm 5 {~\rm
MeV}$, and $f_{D_s^+}/{f_{D^+}=1.24\pm 0.10\pm 0.03}$.
\end{abstract}

\section{Introduction}
To extract precise information from $B-\overline{B}$ mixing
measurements the ratio of ``leptonic decay constants," $f_i$ for
$B_d$ and $B_s$ mesons must be well known \cite{formula-mix}.
Indeed, measurement of $B_s^0-\overline{B}_s^0$ mixing by CDF
\cite{CDF} has pointed out the urgent need for precise numbers. The
$f_i$ have been calculated theoretically. The most promising of
these calculations are based on lattice-gauge theory that include
the light quark loops \cite{Davies}. In order to ensure that these
theories can adequately predict $f_{B_s}/f_{B_d}$ it is critical to
check the analogous ratio from charm decays $f_{D^+_s}/f_{D^+}$. We
have previously measured $f_{D^+}$ \cite{our-fDp,DptomunPRD}.

In the Standard Model (SM) the $D_s$ meson decays purely
leptonically, via annihilation through a virtual $W^+$. The decay
width is given by \cite{Formula1}
\begin{equation}
\Gamma(D_s^+\to \ell^+\nu)  {{G_F^2}\over
8\pi}f_{D_s^+}^2m_{\ell^+}^2M_{D_s^+}\\ [4pt]\label{eq:equ_rate}
\times\left(1-{m_{\ell^+}^2\over M_{D_s^+}^2}\right)^2
\left|V_{cs}\right|^2~, \nonumber
\end{equation}
where $m_{\ell^+}$ and $M_{D_s^+}$ are the $\ell^+$ and $D_s^+$
masses,  $|V_{cs}|$ is a CKM element equal to 0.9737, and $G_F$ is
the Fermi constant.

New physics can affect the expected widths; any undiscovered charged
bosons would interfere with the SM $W^+$ \cite{Akeroyd}. These effects may be
difficult to ascertain, since they would simply change the value of
$f_i$ extracted using Eq.~(1). We can, however, measure the ratio of
decay rates to different leptons, and the predictions then are fixed
only by well-known masses. For example, for $\tau^+\nu$ to
$\mu^+\nu$ the predicted ratio is 9.72; any deviation would be a
manifestation of new physics manifestly violating lepton
universality \cite{Hewett}. 

\section{Experimental Method}
In this study data collected in $e^+e^-$ collisions using the
Cornell Electron Storage Ring (CESR) and recorded near 4.170 GeV.
Here the cross-section for $D_s^{*+}D_s^-$+$D_s^{+}D_s^{*-}$ is
$\sim$1 nb. We fully reconstruct one $D_s$ as a ``tag," and examine
the properties of the other. In this paper we designate the tag as a
$D_s^-$ and examine the leptonic decays of the $D_s^+$, though in
reality we use both charges. For studies with $D_s^+$ decaying into
a $\mu^+\nu$ or $\tau^+\nu$;  $\tau\to\pi^+\overline{\nu}\nu$
($\pi^+\overline{\nu}\nu$) we use 314 pb$^{-1}$ of data; for
$\tau^+\to e^+\overline{\nu}\nu$ ($e^+\overline{\nu}\nu$) we use 195
pb $^{-1}$.

The analysis for $\mu^+\nu$ and $\pi^+\overline{\nu}\nu$ has already been published \cite{Dspubs}.
In summary tags are created from several modes including  $K^+K^-\pi^- $ (13871 events),
$K_S^0 K^-$ (3122), $\eta\pi^-$ (1609), $\eta'\pi^-$ (1196),
$\phi\rho^-$ (1678), $\pi^+\pi^-\pi^-$ (3654), $K^{*-}K^{*0}$ (2030)
and $\eta\rho^-$ (4142), a total of 31302 tags. When the tagging
$\gamma$ from the $D^*$ decay is also required, the number of tags
is reduced to 18645.

Candidate $D_s^+\to\mu^+\nu$ events are searched for by selecting
events with only a single extra track with opposite sign of charge
to the tag; we also require that there not be an extra neutral
energy cluster in excess of 300 MeV. Since here we are searching for
events where there is a single missing neutrino, the missing mass
squared, MM$^2$, evaluated by taking into account the seen $\mu^+$,
$D_s^-$, and the $\gamma$ should peak at zero, and is given by
\begin{equation}
\label{eq:mm2} {\rm MM}^2 = \left(E_{\rm
CM}-E_{D}-E_{\gamma}-E_{\mu}\right)^2\\\nonumber
           -\left(-\overrightarrow{p_
         D}-\overrightarrow{p_{\gamma}}
           -\overrightarrow{p_{\mu}}\right)^2,
\end{equation}
where $E_{\mu}$ ($\overrightarrow{p_{\mu}}$) is the energy
(momentum) of the candidate muon track.

We also make use of a set of kinematical constraints and fit the
MM$^2$ for each $\gamma$ candidate to two hypotheses one of which is
that the $D_s^-$ tag is the daughter of a $D_s^{*-}$ and the other
that the $D_s^{*+}$ decays into $\gamma D_s^+$, with the $D_s^+$
subsequently decaying into $\mu^+\nu$. The hypothesis with the
lowest $\chi^2$ is kept. If there is more than one $\gamma$
candidate in an event we choose only the lowest $\chi^2$ choice
among all the candidates and hypotheses.

The kinematical constraints are the total momentum and energy, the
energy of the either the $D_s^*$ or the $D_s$, the appropriate
$D_s^* - D_s$ mass difference and the invariant mass of the $D_s$
tag decay products.
 This gives us a total of 7 constraints. The
missing neutrino four-vector needs to be determined, so we are left
with a three-constraint fit. We preform a standard iterative fit
minimizing $\chi^2$. As we do not want to be subject to systematic
uncertainties that depend on understanding the absolute scale of the
errors, we do not make a $\chi^2$ cut, but simply choose the photon
and the decay sequence in each event with the minimum $\chi^2$.

We consider three separate cases: (i) the track deposits $<$~300 MeV
in the calorimeter, characteristic of a non-interacting $\pi^+$ or a
$\mu^+$; (ii) the track deposits $>$~300 MeV in the calorimeter,
characteristic of an interacting $\pi^+$; (iii) the track satisfies
our $e^+$ selection criteria \cite{our-fDp}. Then we separately
study the MM$^2$ distributions for these three cases. The separation
between $\mu^+$ and $\pi^+$ is not unique. Case (i) contains 99\% of
the $\mu^+$ but also 60\% of the $\pi^+$, while case (ii) includes
1\% of the $\mu^+$ and 40\% of the $\pi^+$ \cite {DptomunPRD}.

\begin{figure}[h]
\begin{minipage}{18pc}
\includegraphics[width=15pc]{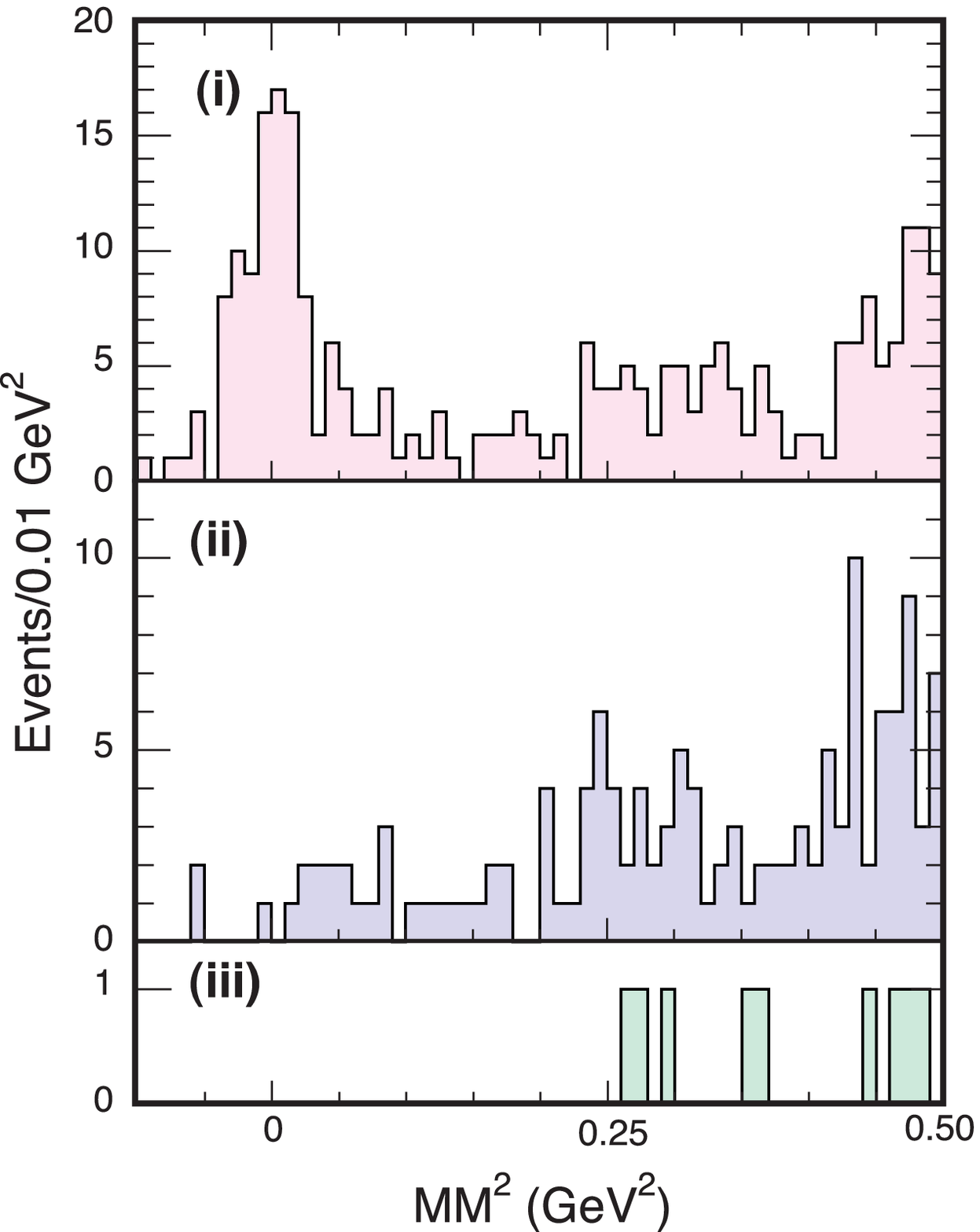}
\caption{\label{mm2-data}The MM$^2$ distributions from data using
$D_s^-$ tags and
 one additional opposite-sign
charged track and no extra energetic showers for cases i-iii (see
text).}
\end{minipage}\hspace{2pc}%
\begin{minipage}{18pc}\vspace*{-4mm}
\includegraphics[width=16pc]{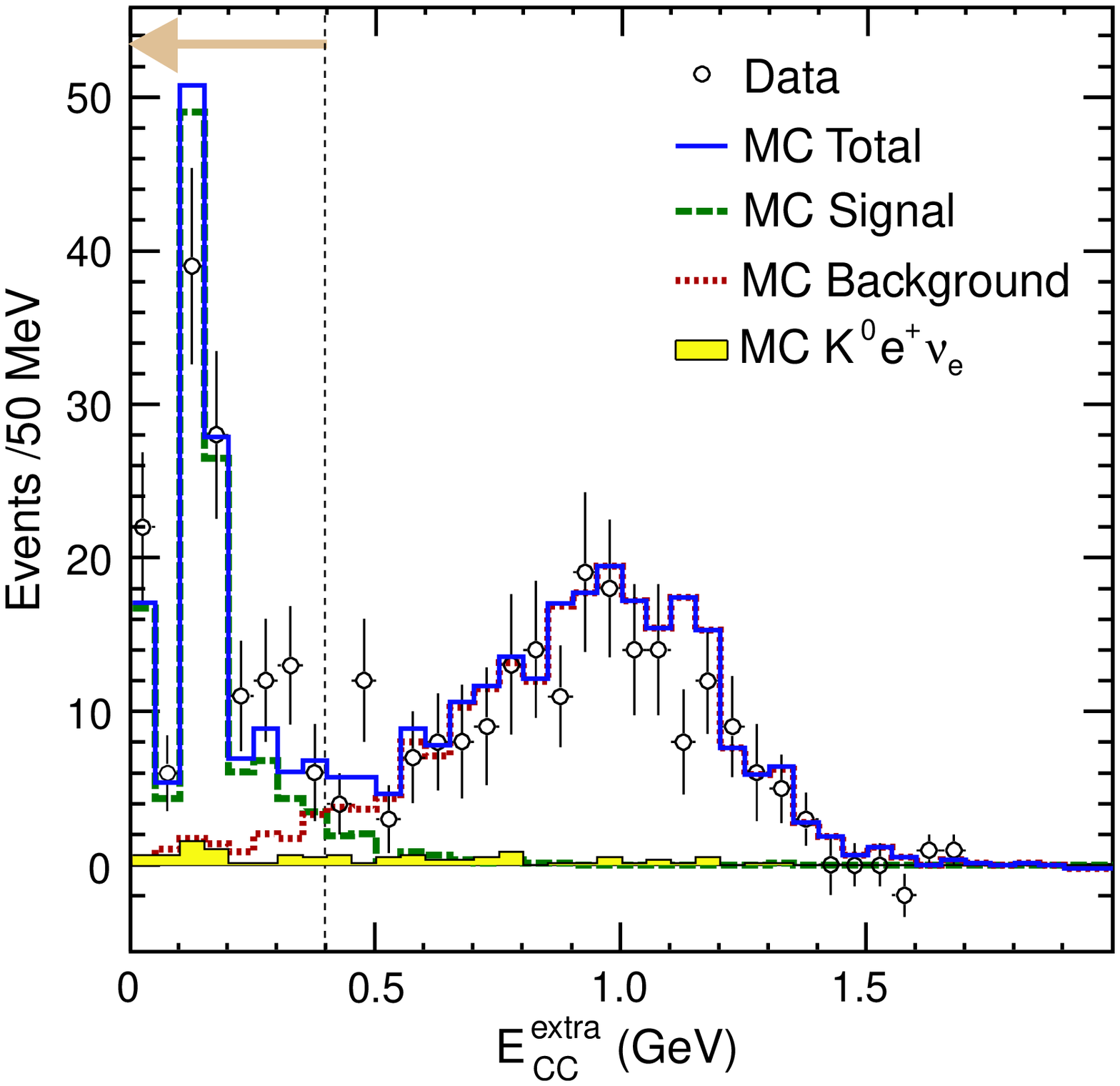}
\caption{\label{ecc-8}The extra calorimeter energy from data
(points), compared with
 the Monte Carlo simulated estimates of semileptonic decays in general (dotted),
 the $K^0 e^+\nu$ mode specifically (shaded),
 a subset of the semileptonics, and the expectation from signal (dashed).
The peak near 150 MeV is due to the $\gamma$ from $D_s^*\to\gamma
D_s$ decay. (The sum is also shown (line).) The arrow indicates the
selected signal region below 0.4 GeV.}
\end{minipage}
\end{figure}

The overall signal region we consider is below MM$^2$ of 0.20
GeV$^2$. Otherwise we admit background from $\eta\pi^+$ and
$K^0\pi^+$ final states. There is a clear peak in
Fig.~\ref{mm2-data}(i), due to $D_s^+\to\mu^+\nu$. Furthermore, the
events in the region between $\mu^+\nu$ peak and 0.20 GeV$^2$ are
dominantly due to the $\tau^+\nu$ decay.

The specific signal regions are defined as follows: for $\mu^+\nu$,
$0.05>{\rm MM}^2>-0.05$ GeV$^2$, corresponding to $\pm 2\sigma$ or 95\%
of the signal; for $\tau\nu$, $\tau^+\to\pi^+\overline{\nu}$, in
case (i) $0.20>{\rm MM}^2>0.05$ GeV$^2$ and in case (ii)
$0.20>{\rm MM}^2>-0.05$ GeV$^2$. In these regions we find 64, 24 and 12
events, respectively. The corresponding backgrounds are estimated as
1, 2.5 and 1 event, respectively.

The $D_s^+\to\tau^+\nu$, $\tau^+\to e^+\nu \overline{\nu}$ mode is
measured by detecting electrons of opposite sign to the tag in
events without any additional charged tracks, and determining the
unmatched energy in the crystal calorimeter (${\rm
E^{extra}_{CC}}$). This energy distribution is shown in
Fig.~\ref{ecc-8}. Requiring ${\rm E^{extra}_{CC}<}$ 400 MeV,
enhances the signal.

 We find
${\cal{B}}(D_s^+\to\mu^+\nu)=(0.594\pm 0.066\pm0.031$)\%; adding in
the $\pi^+\overline{\nu}\nu$ gives ($0.638\pm 0.059\pm0.033$)\% as
an effective rate. Our two measurements for $D_s^+\to\tau^+\nu$ are
($8.0\pm 1.3\pm0.4$)\% and ($6.29\pm 0.78\pm0.52$)\% in the $\pi^+$
and $e^+$ modes, respectively. Finally ${\cal{B}}(D_s^+\to e^+\nu)<
1.3\times 10^{-4}$ (at 90\% cl).

\section{Conclusions}
 We measure
${\Gamma(D_s^+\to \tau^+\nu)}/{\Gamma(D_s^+\to \mu^+\nu)}= 11.5\pm
1.9$, consistent with the SM expectation of 9.72. Combining all three
branching ratios determinations and using $\tau_{D_s^+}$=0.50 ps to
find the leptonic width, we find $f_{D_s}=275\pm 10 \pm 5 {~\rm
MeV}.$ Using our previous result \cite{our-fDp} $f_{D}^+=222.6\pm
16.7^{+2.8}_{-3.4}{\rm ~MeV,}$ provides a determination of
$\displaystyle{{f_{D_s^+}}/{f_{D^+}}=1.24\pm 0.10\pm 0.03}.$

These results are consistent with most recent theoretical models.
The most accurate unquenched lattice model of Follana \etal
\cite{Follana} predicts $f_{D_s}/f_{D^+}$=
$1.162\pm0.009$~\cite{others}.

\section*{Acknowledgments}
This work was supported by the National Science Foundation. I thank
Nabil Menaa for essential discussions.

\end{document}